\documentclass[useAMS,usegraphicx,usenatbib]{mn2e}          

\usepackage{times}

\newcommand{\Msun}{\ensuremath{\,{\rm M}_\odot}}                  
\newcommand{\Rsun}{\ensuremath{\,{\rm R}_\odot}}                  
\newcommand{\Teff}{\ensuremath{T_{\rm eff}}}                      
\newcommand{\logg}{\ensuremath{\log g}}                           
\newcommand{\Mjup}{\ensuremath{\,{\rm M}_{\rm Jup}}}              
\newcommand{\Rjup}{\ensuremath{\,{\rm R}_{\rm Jup}}}              
\newcommand{\Teq}{\ensuremath{T_{\rm eq}^{\,\prime}}}             
\newcommand{\safronov}{\ensuremath{\Theta}}                       
\newcommand{\kms}{\,km\,s$^{-1}$}                                 
\newcommand{\ms}{\,m\,s$^{-1}$}                                   
\newcommand{\mss}{\,m\,s$^{-2}$}                                  
\newcommand{\as}{\ensuremath{^{\prime\prime}}}                    
\newcommand{\am}{\ensuremath{^\prime}}                            
\newcommand{\FeH}{\ensuremath{\left[\frac{\rm Fe}{\rm H}\right]}} 
\newcommand{\pjup}{\ensuremath{\,\rho_{\rm Jup}}}                 
\newcommand{\psun}{\ensuremath{\,\rho_\odot}}                     
\newcommand{\chir}{\ensuremath{\chi_\nu^{\,2}}}                   
\newcommand{\mc}[1]{\multicolumn{2}{c}{#1}}
\newcommand{\mcc}[1]{\multicolumn{3}{c}{#1}}
\newcommand{\er}[3]{\ensuremath{#1^{+#2}_{-#3}}}
\newcommand{\erc}[3]{\mc{\ensuremath{#1^{+#2}_{-#3}}}}

\newcommand{\ermcc}[5]{\mcc{\ensuremath{{#1\,^{+#2}_{-#3}}\,^{+#4}_{-#5}}}}
\newcommand{\reff}[1]{{#1}}

\title[The ultra-short period planet WASP-103]
      {High-precision photometry by telescope defocussing. VII. The ultra-short period planet WASP-103%
      \thanks{Based on data collected by MiNDSTEp with the Danish 1.54\,m telescope, and data collected with GROND on the MPG 2.2\,m telescope, both located at ESO La Silla.}}

\author[Southworth et al.]
       {John Southworth\,$^{1}$,
        L.\ Mancini\,$^{2}$,
        S.\ Ciceri\,$^{2}$,
        J.\ Budaj\,$^{3,4}$,
        M.\ Dominik\,$^{5}$\thanks{Royal Society University Research Fellow},
        \newauthor
        R.\ Figuera Jaimes$^{5,6}$,
        T.\ Haugb{\o}lle\,$^{7}$,
        U.\ G.\ J{\o}rgensen\,$^{8}$,
        A.\ Popovas\,$^{8}$,
        M.\ Rabus\,$^{9,2}$,
        \newauthor
        S.\ Rahvar\,$^{10}$,
        C.\ von Essen\,$^{11}$,
        R.\ W.\ Schmidt\,$^{12}$,
        O.\ Wertz\,$^{13}$,
        K.\ A.\ Alsubai\,$^{14}$,
        \newauthor
        V.\ Bozza\,$^{15,16}$,
        D.\ M.\ Bramich\,$^{14}$,
        S.\ Calchi Novati\thanks{Sagan visiting fellow}\,$^{17,15,18}$,
        G.\ D'Ago\,$^{15,16}$,
        \newauthor
        T.\ C.\ Hinse\,$^{19}$,
        Th.\ Henning\,$^{2}$,
        M.\ Hundertmark\,$^{5}$,
        D.\ Juncher\,$^{8}$,
        H.\ Korhonen\,$^{20,8}$,
        \newauthor
        J.\ Skottfelt\,$^{8}$,
        C.\ Snodgrass\,$^{21}$,
        D.\ Starkey\,$^{5}$
        J.\ Surdej\,$^{13}$
        \\
        $^{1}$\,Astrophysics Group, Keele University, Staffordshire, ST5 5BG, UK \\
        $^{2}$\,Max Planck Institute for Astronomy, K\"onigstuhl 17, 69117 Heidelberg, Germany \\
        $^{3}$\,Research School of Astronomy and Astrophysics, Australian National University, Canberra, ACT 2611, Australia \\
        $^{4}$\,Astronomical Institute of the Slovak Academy of Sciences, 059 60 Tatransk\'a Lomnica, Slovakia \\
        $^{5}$\,SUPA, University of St Andrews, School of Physics \& Astronomy, North Haugh, St Andrews, KY16 9SS, UK \\
        $^{6}$\,European Southern Observatory, Karl-Schwarzschild-Stra{\ss}e 2, 85748 Garching bei M\"unchen, Germany \\
        $^{7}$\,Centre for Star and Planet Formation, Natural History Museum, University of Copenhagen, {\O}ster Voldgade 5-7, 1350 Copenhagen K, Denmark \\
        $^{8}$\,Niels Bohr Institute \& Centre for Star and Planet Formation, University of Copenhagen, Juliane Maries vej 30, 2100 Copenhagen \O, Denmark \\
        $^{9}$\,Instituto de Astrof{\i}sica, Facultad de F{\i}sica, Pontificia Universidad Cat\'olica de Chile, Av.\ Vicu\~na Mackenna 4860, 7820436 Macul, Santiago, Chile \\
        $^{10}$\,Department of Physics, Sharif University of Technology, P.\,O.\,Box 11155-9161 Tehran, Iran \\
        $^{11}$\,Stellar Astrophysics Centre, Department of Physics and Astronomy, Aarhus University, Ny Munkegade 120, DK-8000 Aarhus C, Denmark \\
        $^{12}$\,Astronomisches Rechen-Institut, Zentrum f\"ur Astronomie, Universit\"at Heidelberg, M\"onchhofstra{\ss}e 12-14, 69120 Heidelberg, Germany \\
        $^{13}$\,Institut d'Astrophysique et de G\'eophysique, Universit\'e de Li\`ege, 4000 Li\`ege, Belgium \\
        $^{14}$\,Qatar Environment and Energy Research Institute, Qatar Foundation, Tornado Tower, Floor 19, P.O.\ Box 5825, Doha, Qatar \\
        $^{15}$\,Dipartimento di Fisica ``E.R. Caianiello'', Universit\`a di Salerno, Via Giovanni Paolo II 132, 84084, Fisciano (SA), Italy \\
        $^{16}$\,Istituto Nazionale di Fisica Nucleare, Sezione di Napoli, Napoli, Italy \\
        $^{17}$\,NASA Exoplanet Science Institute, MS 100-22, California Institute of Technology, Pasadena, CA 91125, US \\
        $^{18}$\,Istituto Internazionale per gli Alti Studi Scientifici (IIASS), 84019 Vietri Sul Mare (SA), Italy \\
        $^{19}$\,Korea Astronomy and Space Science Institute, Daejeon 305-348, Republic of Korea \\
        $^{20}$\,Finnish Centre for Astronomy with ESO (FINCA), University of Turku, V{\"a}is{\"a}l{\"a}ntie 20, FI-21500 Piikki{\"o}, Finland \\
        $^{21}$\,Planetary and Space Sciences, Department of Physical Sciences, The Open University, Milton Keynes, MK7 6AA, UK \\
        }

\begin{document} \maketitle 

\clearpage

\begin{abstract}
We present 17 transit light curves of the ultra-short period planetary system WASP-103, a strong candidate for the detection of tidally-induced orbital decay. We use these to establish a high-precision reference epoch for transit timing studies. The time of the reference transit midpoint is now measured to an accuracy of 4.8\,s, versus 67.4\,s in the discovery paper, aiding future searches for orbital decay. With the help of published spectroscopic measurements and theoretical stellar models, we determine the physical properties of the system to high precision and present a detailed error budget for these calculations. The planet \reff{has a Roche lobe filling factor of 0.58}, leading to a significant asphericity; we correct its measured mass and mean density for this phenomenon. A high-resolution {\it Lucky Imaging} observation shows no evidence for faint stars close enough to contaminate the point spread function of WASP-103. Our data were obtained in the Bessell $RI$ and the SDSS $griz$ passbands and yield a larger planet radius at bluer optical wavelengths, to a confidence level of $7.3\sigma$. Interpreting this as an effect of Rayleigh scattering in the planetary atmosphere leads to a measurement of the planetary mass which is too small by a factor of five, implying that Rayleigh scattering is not the main cause of the variation of radius with wavelength.
\end{abstract}



\section{Introduction}                                                                                                              \label{sec:intro}

An important factor governing the tidal evolution of planetary systems is the stellar tidal quality factor $Q_\star$ \citep[e.g.][]{GoldreichSoter66icar}, which represents the efficiency of tidal dissipation in the star. Its value is necessary for predicting the timescales of orbital circularisation, axial alignment and rotational synchronisation of binary star and planet systems. Short-period giant planets suffer orbital decay due to tidal effects, and most will ultimately be devoured by their host star rather than reach an equilibrium state \citep{Levrard++09apj,Jackson++09apj}. The magnitude of $Q_\star$ therefore influences the orbital period distribution of populations of extrasolar planets.

Unfortunately, $Q_\star$ is not well constrained by current observations. Its value is often taken to be $10^6$ \citep{OgilvieLin07apj} but there exist divergent results in the literature. A value of $10^{5.5}$ was found to be a good match to a sample of known extrasolar planets by \citet{Jackson++08apj2}, but theoretical work by \citet{PenevSasselov11apj} constrained $Q_\star$ to lie between $10^8$ and $10^{9.5}$ and an observational study by \citet{Penev+12apj} found $Q_\star > 10^7$ to 99\% confidence. Inferences from the properties of binary star systems are often used but are not relevant to this issue: $Q_\star$ is not a fundamental property of a star but depends on the nature of the tidal perturbation \citep{Goldreich63mn,Ogilvie14xxx}. $Q_\star$ should, however, be observationally accessible through the study of transiting extrasolar planets (TEPs).

\citet{Birkby+14mn} assessed the known population of TEPs for their potential for the direct determination of the strength of tidal interations. The mechanism considered was the detection of tidally-induced orbital decay, which manifests itself as a decreasing orbital period. These authors found that WASP-18 \citep{Hellier+09nat,Me+09apj} is the most promising system, due to its short orbital period (0.94\,d) and large planet mass (10.4\Mjup), followed by WASP-103 \citep[][hereafter G14]{Gillon+14aa}, then WASP-19 \citep{Hebb+10apj,Mancini+13mn}.

Adopting the canonical value of $Q_\star = 10^6$, \citet{Birkby+14mn} calculated that orbital decay would cause a shift in transit times -- over a time interval of 10\,yr -- of 350\,s for WASP-18, 100\,s for WASP-103 and 60\,s for WASP-19. Detection of this effect clearly requires observations over many years coupled with a precise ephemeris against which to measure deviations from strict periodicity. High-quality transit timing data are already available for WASP-18 \citep{Maxted+13mn} and WASP-19 \citep{Tregloan++13mn,Abe+13aa,Lendl+13aa,Mancini+13mn}, but not for WASP-103.

WASP-103 was discovered by G14 and comprises a TEP of mass 1.5\Mjup\ and radius 1.6\Rjup\ in a very short-period orbit (0.92\,d) around an F8\,V star of mass 1.2\Msun\ and radius 1.4\Rsun. G14 obtained observations of five transits, two with the Swiss Euler telescope and three with TRAPPIST, both at ESO La Silla. The Euler data each cover only half a transit, whereas the TRAPPIST data have a lower photometric precision and suffer from 180\degr\ field rotations during the transits due to the nature of the telescope mount. The properties of the system could therefore be measured to only modest precision; in particular the ephemeris zeropoint is known to a precision of only 64\,s. In this work we present 17 high-quality transit light curves which we use to determine a precise orbital ephemeris for WASP-103, as well as to improve measurements of its physical properties.


\section{Observations and data reduction}                                                                                             \label{sec:obs}

\begin{table*} \centering
\caption{\label{tab:obslog} Log of the observations presented in this work. $N_{\rm obs}$ is the number
of observations, $T_{\rm exp}$ is the exposure time, $T_{\rm dead}$ is the dead time between exposures,
`Moon illum.' is the fractional illumination of the Moon at the midpoint of the transit, and $N_{\rm poly}$
is the order of the polynomial fitted to the out-of-transit data. The aperture radii are target aperture,
inner sky and outer sky, respectively.}
\setlength{\tabcolsep}{5pt}
\begin{tabular}{lcccccccccccc} \hline
Instrument & Date of   & Start time & End time  &$N_{\rm obs}$ & $T_{\rm exp}$ & $T_{\rm dead}$ & Filter & Airmass & Moon & Aperture   & $N_{\rm poly}$ & Scatter \\
           & first obs &    (UT)    &   (UT)    &              & (s)           & (s)            &        &         &illum.& radii (px) &                & (mmag)  \\
\hline
DFOSC       & 2014 04 20 & 05:08 & 09:45 & 134 & 100--105 & 18 & $R$ & 1.54 $\to$ 1.24 $\to$ 1.54 & 0.725 & 14 25 45 & 1 & 0.675 \\ 
DFOSC       & 2014 05 02 & 05:45 & 10:13 & 113 & 110--130 & 19 & $I$ & 1.28 $\to$ 1.24 $\to$ 2.23 & 0.100 & 14 22 50 & 1 & 0.815 \\ 
DFOSC       & 2014 06 09 & 04:08 & 08:22 & 130 & 100      & 16 & $R$ & 1.24 $\to$ 3.08            & 0.888 & 16 27 50 & 1 & 1.031 \\ 
DFOSC       & 2014 06 23 & 01:44 & 06:19 & 195 & 50--120  & 16 & $R$ & 1.35 $\to$ 1.24 $\to$ 1.86 & 0.168 & 14 22 40 & 1 & 1.329 \\ 
DFOSC       & 2014 06 24 & 00:50 & 04:25 & 112 & 100      & 16 & $R$ & 1.54 $\to$ 1.24 $\to$ 1.31 & 0.103 & 19 25 40 & 1 & 0.647 \\ 
DFOSC       & 2014 07 06 & 01:28 & 05:20 & 118 & 100      & 18 & $R$ & 1.28 $\to$ 1.24 $\to$ 1.80 & 0.564 & 16 26 50 & 1 & 0.653 \\ 
DFOSC       & 2014 07 18 & 01:19 & 05:45 & 139 & 90--110  & 16 & $R$ & 1.25 $\to$ 3.00            & 0.603 & 17 25 60 & 1 & 0.716 \\ 
DFOSC       & 2014 07 18 & 23:04 & 04:20 & 181 & 60--110  & 16 & $R$ & 1.52 $\to$ 1.24 $\to$ 1.73 & 0.502 & 16 24 55 & 2 & 0.585 \\[3pt] 
GROND       & 2014 07 06 & 00:23 & 05:27 & 122 & 100--120 & 40 & $g$ & 1.45 $\to$ 1.24 $\to$ 1.87 & 0.564 & 24 65 85 & 2 & 1.251 \\ 
GROND       & 2014 07 06 & 00:23 & 05:27 & 119 & 100--120 & 40 & $r$ & 1.45 $\to$ 1.24 $\to$ 1.87 & 0.564 & 24 65 85 & 2 & 0.707 \\ 
GROND       & 2014 07 06 & 00:23 & 05:27 & 125 & 100--120 & 40 & $i$ & 1.45 $\to$ 1.24 $\to$ 1.87 & 0.564 & 24 65 85 & 2 & 0.843 \\ 
GROND       & 2014 07 06 & 00:23 & 05:27 & 121 & 100--120 & 40 & $z$ & 1.45 $\to$ 1.24 $\to$ 1.87 & 0.564 & 24 65 85 & 2 & 1.106 \\ 
GROND       & 2014 07 18 & 22:55 & 03:59 & 125 & 98--108  & 41 & $g$ & 1.64 $\to$ 1.24 $\to$ 1.93 & 0.502 & 30 50 85 & 2 & 0.882 \\ 
GROND       & 2014 07 18 & 22:55 & 04:43 & 143 & 98--108  & 41 & $r$ & 1.64 $\to$ 1.24 $\to$ 1.93 & 0.502 & 25 45 70 & 2 & 0.915 \\ 
GROND       & 2014 07 18 & 22:55 & 04:39 & 142 & 98--108  & 41 & $i$ & 1.64 $\to$ 1.24 $\to$ 1.88 & 0.502 & 28 56 83 & 2 & 0.656 \\ 
GROND       & 2014 07 18 & 22:55 & 04:43 & 144 & 98--108  & 41 & $z$ & 1.64 $\to$ 1.24 $\to$ 1.93 & 0.502 & 30 50 80 & 2 & 0.948 \\[3pt] 
CASLEO      & 2014 08 12 & 23:22 & 03:10 & 129 & 90--120  &  4 & $R$ & 1.29 $\to$ 2.12            & 0.920 & 20 30 60 & 4 & 1.552 \\ 
\hline \end{tabular} \end{table*}                                                                                                   

\begin{figure} \includegraphics[width=\columnwidth,angle=0]{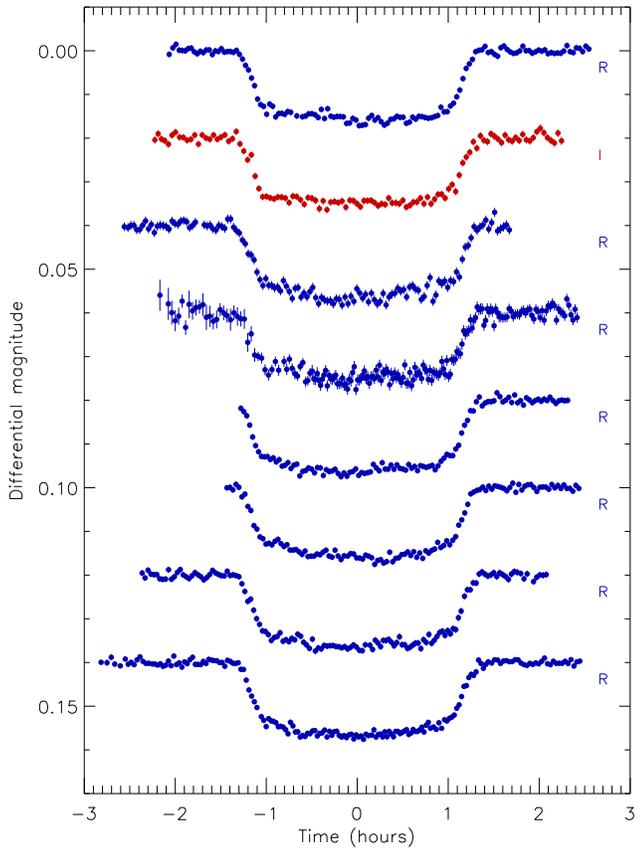}
\caption{\label{fig:lc:dk} DFOSC light curves presented in this work, in the
order they are given in Table\,\ref{tab:obslog}. Times are given relative to
the midpoint of each transit, and the filter used is indicated. Dark blue and
dark red filled circles represent observations through the Bessell $R$ and
$I$ filters, respectively.} \end{figure}

\begin{figure} \includegraphics[width=\columnwidth,angle=0]{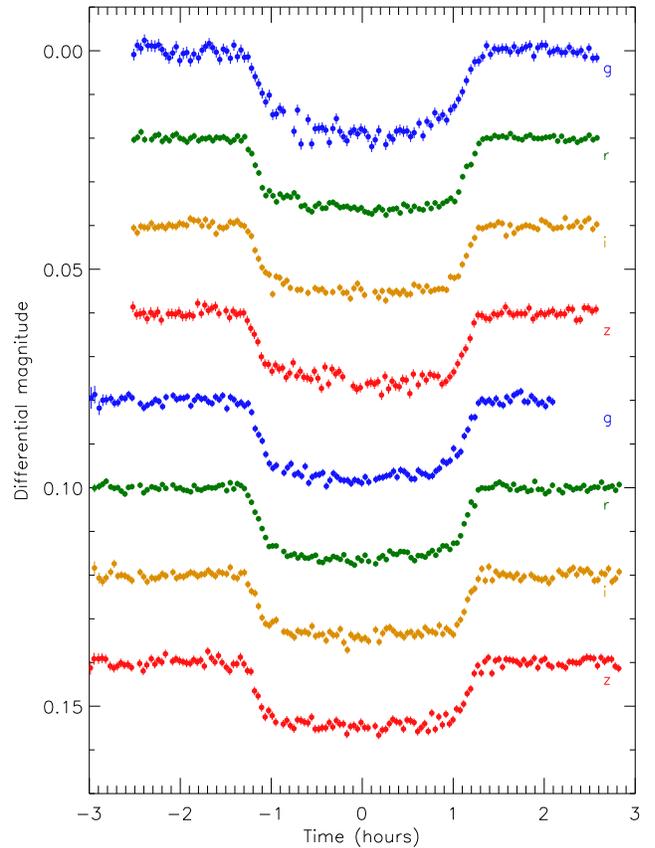}
\caption{\label{fig:lc:grond} GROND light curves presented in this work, in the order
they are given in Table\,\ref{tab:obslog}. Times are given relative to the midpoint
of each transit, and the filter used is indicated. $g$-band data are shown in light
blue, $r$-band in green, $i$-band in orange and $z$-band in light red.} \end{figure}

\begin{figure} \includegraphics[width=\columnwidth,angle=0]{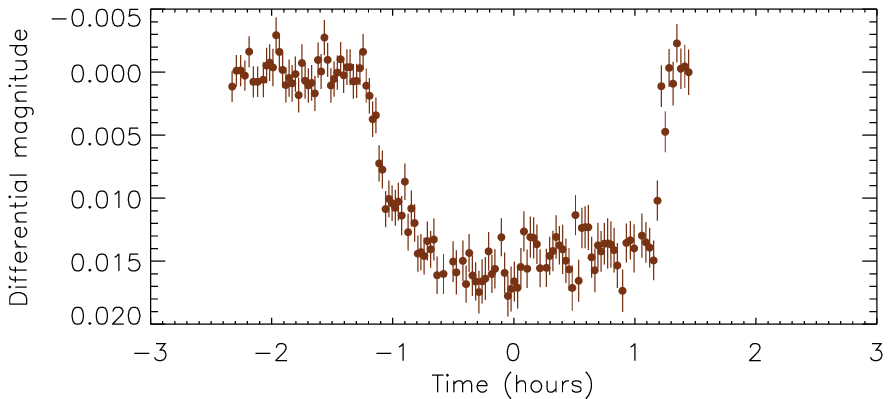}
\caption{\label{fig:lc:casleo} The CASLEO light curve of WASP-103. Times are given
relative to the midpoint of the transit.} \end{figure}

\subsection{DFOSC observations}                                                                                                 \label{sec:obs:dfosc}

Eight transits were obtained using the DFOSC (Danish Faint Object Spectrograph and Camera) instrument on the 1.54\,m Danish Telescope at ESO La Silla, Chile, in the context of the MiNDSTEp microlensing program \citep{Dominik+10an}. DFOSC has a field of view of 13.7\am$\times$13.7\am\ at a plate scale of 0.39\as\,pixel$^{-1}$. We windowed down the CCD to cover WASP-103 itself and seven good comparison stars, in order to shorten the dead time between exposures.

The instrument was defocussed to lower the noise level of the observations, in line with our usual strategy \citep[see][]{Me+09mn,Me+14mn}. The telescope was autoguided to limit pointing drifts to less than five pixels over each observing sequence. Seven of the transits were obtained through a Bessell $R$ filter, but one was taken through a Bessell $I$ filter by accident. An observing log is given in Table\,\ref{tab:obslog} and the light curves are plotted individually in Fig.\,\ref{fig:lc:dk}.

\subsection{GROND observations}                                                                                                 \label{sec:obs:grond}

We observed two transits of WASP-103 using the GROND instrument \citep{Greiner+08pasp} mounted on the MPG 2.2\,m telescope at La Silla, Chile. Both transits were also observed with DFOSC. GROND was used to obtain light curves simultaneously in passbands which approximate SDSS $g$, $r$, $i$ and $z$. The small field of view of this instrument (5.4$^{\prime}$$\times$5.4$^{\prime}$ at a plate scale of 0.158$^{\prime\prime}$\,pixel$^{-1}$) meant that few comparison stars were available and the best of these was several times fainter than WASP-103 itself. The scatter in the GROND light curves is therefore worse than generally achieved \citep[e.g.][]{Nikolov+13aa,Mancini+14aa,Mancini+14aa2}, but the data are certainly still useful. The telescope was defocussed and autoguided for both sets of observations. Further details are given in the observing log (Table\,\ref{tab:obslog}) and the light curves are plotted individually in Fig.\,\ref{fig:lc:grond}.

\subsection{CASLEO observations}

We observed one transit of WASP-103 (Fig.\,\ref{fig:lc:casleo}) using the 2.15\,m Jorge Sahade telescope located at the Complejo Astron\'omico El Leoncito in San Juan, Argentina\footnote{Visiting Astronomer, Complejo Astron\'omico El Leoncito operated under agreement between the Consejo Nacional de Investigaciones Cient\'{\i}ficas y T\'ecnicas de la Rep\'ublica Argentina and the National Universities of La Plata, C\'ordoba and San Juan}. We used the focal reducer and Roper Scientific CCD, yielding an unvignetted field of view of 9$^{\prime}$ radius at a plate scale of 0.45$^{\prime\prime}$\,pixel$^{-1}$. The CCD was operated without binning or windowing due to its short readout time. The observing conditions were excellent. The images were slightly defocussed to a FWHM of 3$^{\prime\prime}$, and were obtained through a Johnson-Cousins Schuler $R$ filter.

\subsection{Data reduction}                                                                                                     \label{sec:obs:defot}

\begin{table} \centering \caption{\label{tab:lcdata} Sample of the
data presented in this work (the first datapoint of each light curve).
The full dataset will be made available at the CDS.}
\begin{tabular}{l c r@{.}l r@{.}l r@{.}l} \hline
instrument & Filter & \mc{BJD(TDB)} & \mc{Diff.\ mag.} & \mc{Uncertainty} \\
\hline
DFOSC  & $R$ & 2456767&719670 &  0&0008211 & 0&0006953 \\
DFOSC  & $I$ & 2456779&746022 &  0&0004141 & 0&0008471 \\
DFOSC  & $R$ & 2456817&679055 &  0&0002629 & 0&0009941 \\
DFOSC  & $R$ & 2456831&578063 & -0&0040473 & 0&0035872 \\
DFOSC  & $R$ & 2456832&540736 &  0&0018135 & 0&0006501 \\
DFOSC  & $R$ & 2456844&566748 &  0&0000224 & 0&0006405 \\
DFOSC  & $R$ & 2456856&559865 & -0&0004751 & 0&0006952 \\
DFOSC  & $R$ & 2456857&466519 & -0&0000459 & 0&0005428 \\
GROND  & $g$ & 2456844&521804 &  0&0008698 & 0&0013610 \\
GROND  & $r$ & 2456844&521804 &  0&0003845 & 0&0007863 \\
GROND  & $i$ & 2456844&521804 &  0&0005408 & 0&0009272 \\
GROND  & $z$ & 2456844&521804 & -0&0013755 & 0&0012409 \\
GROND  & $g$ & 2456857&459724 & -0&0005583 & 0&0024887 \\
GROND  & $r$ & 2456857&461385 &  0&0000852 & 0&0009575 \\
GROND  & $i$ & 2456857&461385 & -0&0016371 & 0&0015169 \\
GROND  & $z$ & 2456857&459724 &  0&0012460 & 0&0015680 \\
CASLEO & $R$ & 2456882&47535  &  0&00114   & 0&00121   \\
\hline \end{tabular} \end{table}

The DFOSC and GROND data were reduced using the {\sc defot} code \citep{Me+09mn} with the improvements discussed by \citet{Me+14mn}. Master bias, dome flat fields and sky flat fields were constructed but not applied, as they were found not to improve the quality of the resulting light curves \citep[see][]{Me+14mn}. Aperture photometry was performed using the {\sc idl\footnote{The acronym {\sc idl} stands for Interactive Data Language and is a trademark of ITT Visual Information Solutions. For further details see: {\tt http://www.ittvis.com/ProductServices/IDL.aspx}.}/astrolib\footnote{The {\sc astrolib} subroutine library is distributed by NASA. For further details see: {\tt http://idlastro.gsfc.nasa.gov/}.}} implementation of {\sc daophot} \citep{Stetson87pasp}. Image motion was tracked by cross-correlating individual images with a reference image.

We obtained photometry on the instrumental system using software apertures of a range of sizes, and retained those which gave light curves with the smallest scatter (Table\,\ref{tab:obslog}). We found that the choice of aperture size does influence the scatter in the final light curve, but does not have a significant effect on the transit shape.

The instrumental magnitudes were then transformed to differential-magnitude light curves normalised to zero magnitude outside transit. The normalisation was enforced with first- or second-order polynomials (see Table\,\ref{tab:obslog}) fitted to the out-of-transit data. The differential magnitudes are relative to a weighted ensemble of typically five (DFOSC) or two to four (GROND) comparison stars. The comparison star weights and polynomial coefficients were simultaneously optimised to minimise the scatter in the out-of-transit data.

The CASLEO data were reduced using standard aperture photometry methods, with the IRAF tasks {\sc ccdproc} and {\sc apphot}. We found that it was necessary to flat-field the data in order to obtain a good light curve. The final light curve was obtained by dividing the flux of WASP-103 by the average flux of three comparison stars. An aperture radius of three times the FWHM was used, as it minimised the scatter in the data.

Finally, the timestamps for the datapoints were converted to the BJD(TDB) timescale \citep{Eastman++10pasp}. We performed manual time checks for several images and have verified that the FITS file timestamps are on the UTC system to within a few seconds. The reduced data are given in Table\,\ref{tab:lcdata} and will be lodged with the CDS\footnote{{\tt http://vizier.u-strasbg.fr/}}.

\subsection{High-resolution imaging}                                                                                               \label{sec:obs:li}

\begin{figure}
\includegraphics[width=\columnwidth,angle=0]{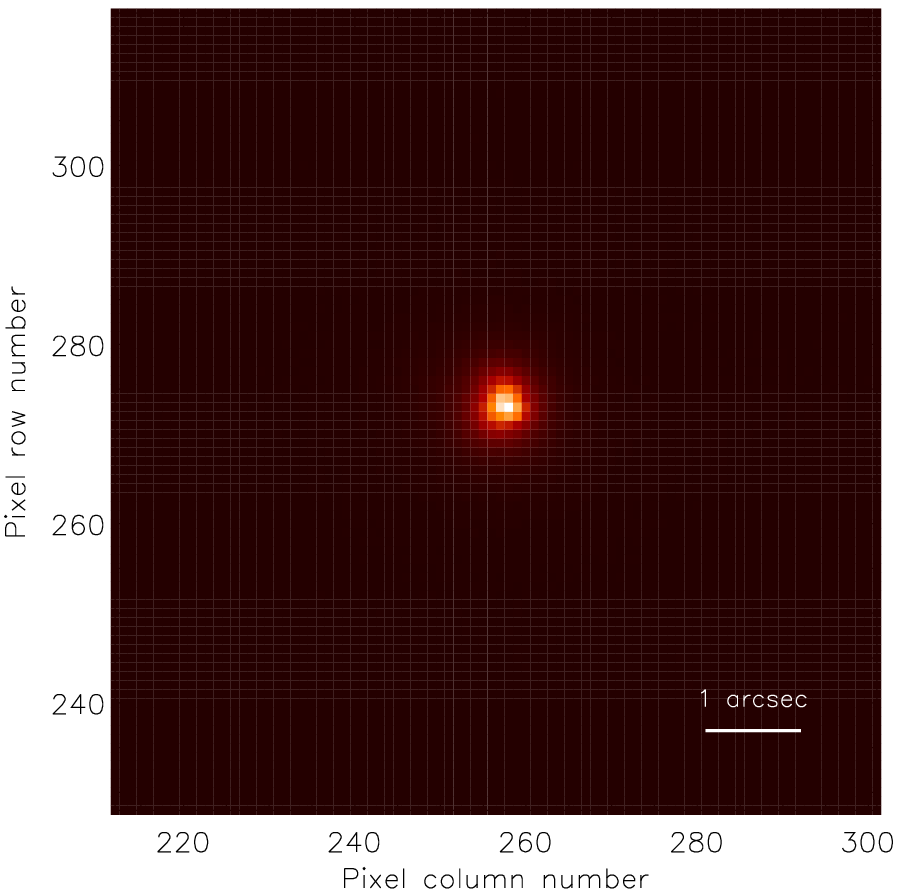}
\includegraphics[width=\columnwidth,angle=0]{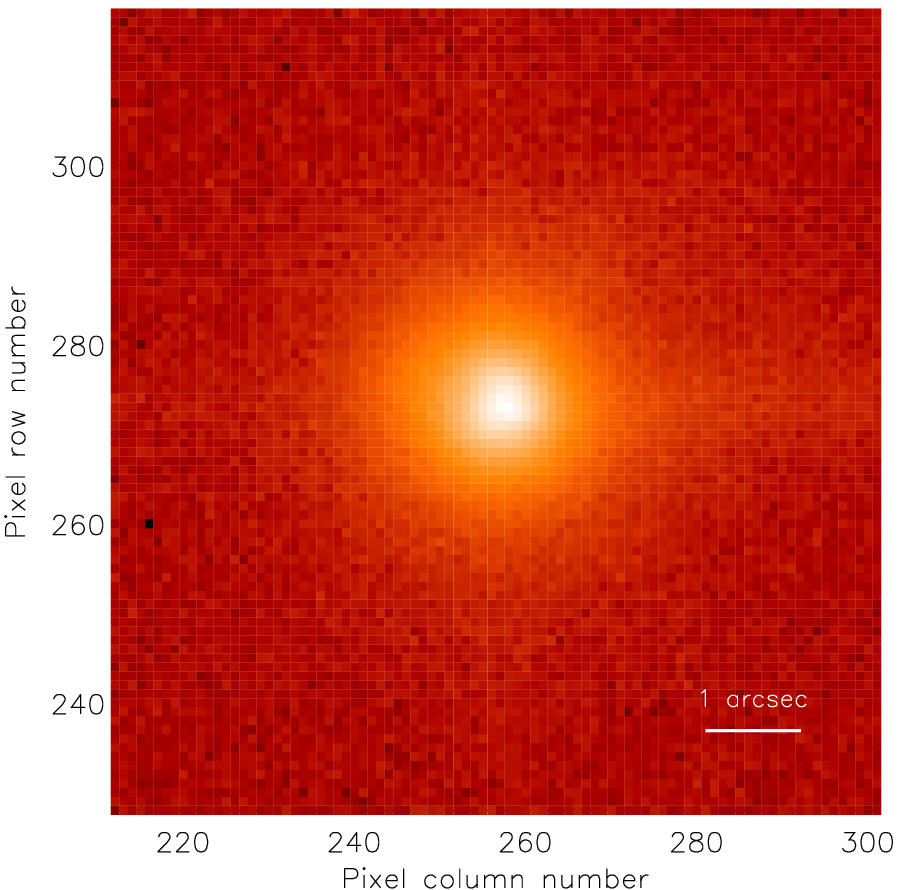}
\caption{\label{fig:li} High-resolution Lucky Image of the field around
WASP-103. The upper panel has a linear flux scale for context and the lower
panel has a logarithmic flux scale to enhance the visibility of any faint
stars. Each image covers $8\as \times 8\as$ centred on WASP-103. A bar of
length $1\as$ is superimposed in the bottom-right of each image. The image
is a sum of the best 2\% of the original images.} \end{figure}

Several images were taken of WASP-103 with DFOSC in sharp focus, in order to test for the presence of faint nearby stars whose photons might bias our results \citep{Daemgen+09aa}. The closest star we found on any image is 42 pixels south-east of WASP-103, and 5.3\,mag fainter in the $R$-band. It is thus too faint and far away to contaminate the inner aperture of our target star.

We proceeded to obtain a high-resolution image of WASP-103 using the Lucky Imager (LI) mounted on the Danish telescope \citep[see][]{Skottfelt+13aa}. The LI uses an Andor 512$\times$512 pixel electron-multiplying CCD, with a pixel scale of 0.09\as\,pixel$^{-1}$ giving a field of view of $45\as\times45\as$. The data were reduced using a dedicated pipeline and the 2\% of images with the smallest point spread function (PSF) were stacked together to yield combined images whose PSF is smaller than the seeing limit. A long-pass filter was used, resulting in a response which approximates that of SDSS $i$$+$$z$. An overall exposure time of 415\,s corresponds to an effective exposure time of 8.3\,s for the best 2\% of the images. The FWHM of the PSF is 5.9\,pixels (0.53\as) in both dimensions. The LI image (Fig.\,\ref{fig:li}) shows no evidence for a point source closer than that found in our DFOSC images.


\section{Transit timing analysis}                                                                                                    \label{sec:porb}

\begin{figure*}
\includegraphics[width=\textwidth,angle=0]{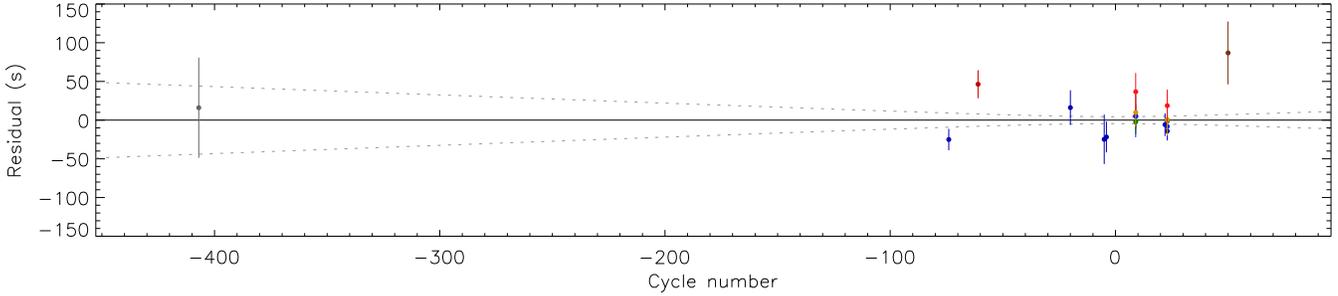}
\caption{\label{fig:minima} Plot of the residuals of the timings of mid-transit
for WASP-103 versus a linear ephemeris. The leftmost point is from G14 and the
remaining points are from the current work (colour-coded consistently with Figs.\
\ref{fig:lc:dk}, \ref{fig:lc:grond} and \ref{fig:lc:casleo}). The dotted lines show the 1$\sigma$
uncertainty in the ephemeris as a function of cycle number.} \end{figure*}

\begin{table} \begin{center}
\caption{\label{tab:minima} Times of minimum light and their residuals versus the ephemeris derived in this work.}
\setlength{\tabcolsep}{3pt}
\begin{tabular}{l l r r l} \hline
Time of min. & Error & Cycle & Residual & Reference \\
(BJD/TDB)  & (d) & number & (d) &       \\
\hline
2456459.59957 & 0.00079 & $-$407.0 &    0.00019 & G14                   \\   
2456767.80578 & 0.00017 &  $-$74.0 & $-$0.00029 & This work (DFOSC $R$) \\   
2456779.83870 & 0.00022 &  $-$61.0 &    0.00054 & This work (DFOSC $I$) \\   
2456817.78572 & 0.00027 &  $-$20.0 &    0.00019 & This work (DFOSC $R$) \\   
2456831.66843 & 0.00039 &   $-$5.0 & $-$0.00029 & This work (DFOSC $R$) \\   
2456832.59401 & 0.00024 &   $-$4.0 & $-$0.00025 & This work (DFOSC $R$) \\   
2456844.62641 & 0.00019 &      9.0 &    0.00005 & This work (DFOSC $R$) \\   
2456844.62642 & 0.00034 &      9.0 &    0.00006 & This work (GROND $g$) \\   
2456844.62633 & 0.00019 &      9.0 & $-$0.00003 & This work (GROND $r$) \\   
2456844.62647 & 0.00023 &      9.0 &    0.00011 & This work (GROND $i$) \\   
2456844.62678 & 0.00030 &      9.0 &    0.00042 & This work (GROND $z$) \\   
2456856.65838 & 0.00018 &     22.0 & $-$0.00007 & This work (DFOSC $R$) \\   
2456857.58383 & 0.00014 &     23.0 & $-$0.00016 & This work (DFOSC $R$) \\   
2456857.58390 & 0.00022 &     23.0 & $-$0.00009 & This work (GROND $g$) \\   
2456857.58398 & 0.00017 &     23.0 &    0.00001 & This work (GROND $r$) \\   
2456857.58400 & 0.00025 &     23.0 &    0.00001 & This work (GROND $i$) \\   
2456857.58421 & 0.00025 &     23.0 &    0.00022 & This work (GROND $z$) \\   
2456882.57473 & 0.00050 &     50.0 &    0.00100 & This work (CASLEO $R$)\\   
\hline \end{tabular} \end{center} \end{table}

We first modelled each light curve individually using the {\sc jktebop} code (see below) in order to determine \reff{the} times of mid-transit. In this process the errorbars for each dataset were also scaled to give a reduced $\chi^2$ of $\chi^2_\nu = 1.0$ versus the fitted model. This step is needed because the uncertainties from the {\sc aper} algorithm are often moderately underestimated.

We then fitted the times of mid-transit with a straight line versus cycle number to determine a new linear orbital ephemeris. We included the ephemeris zeropoint from G14, which is also on the BJD(TDB) timescale and was obtained by them from a joint fit to all their data. Table\,\ref{tab:minima} gives all transit times plus their residual versus the fitted ephemeris. We chose the reference epoch to be that which gives the lowest uncertainty in the time zeropoint, as this minimises the covariance between the reference time of minimum and the orbital period. The resulting ephemeris is
$$ T_0 = {\rm BJD(TDB)} \,\, 2\,456\,836.296445 (55) \, + \, 0.9255456 (13) \times E $$
where $E$ gives the cycle count versus the reference epoch and the bracketed quantities indicate the uncertainty in the final digit of the preceding number.

The $\chi^2_\nu$ of the fit is excellent at 1.055. The timestamps from DFOSC and GROND are obtained from different atomic clocks, so are unrelated to each other. The good agreement between them is therefore evidence that both are correct.

Fig.\,\ref{fig:minima} shows the residuals of the times of mid-transit versus the linear ephemeris we have determined. The precision in the measurement of the midpoint of the reference transit has improved from 64.7\,s (G14) to 4.8\,s, meaning that we have established a high-quality set of timing data against which orbital decay could be measured in future.


\section{Light curve analysis}                                                                                                         \label{sec:lc}

\begin{figure} \includegraphics[width=\columnwidth,angle=0]{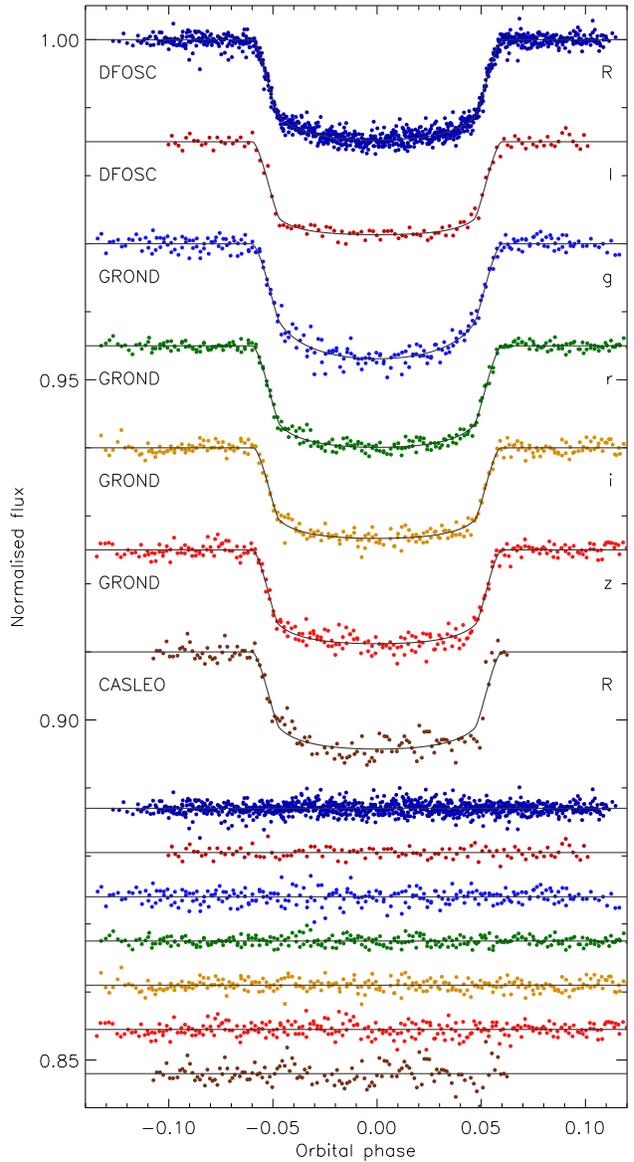}
\caption{\label{fig:lcfit} Phased light curves of WASP-103 compared to the
{\sc jktebop} best fits. The residuals of the fits are plotted at the base
of the figure, offset from unity. Labels give the source and passband for
each dataset. The polynomial baseline functions have been removed from the
data before plotting.} \end{figure}

\begin{table*} \centering \caption{\label{tab:lcfit} Parameters of the
fit to the light curves of WASP-103 from the {\sc jktebop} analysis (top).
The final parameters are given in bold and the parameters found by G14 are
given below this. Quantities without quoted uncertainties were not given by
G14 but have been calculated from other parameters which were. The errorbars
for the final value of $k$ has been inflated to account for the disagreement
between different measurements.}
\begin{tabular}{l r@{\,$\pm$\,}l r@{\,$\pm$\,}l r@{\,$\pm$\,}l r@{\,$\pm$\,}l r@{\,$\pm$\,}l}
\hline
Source       & \mc{$r_{\rm A}+r_{\rm b}$} & \mc{$k$} & \mc{$i$ ($^\circ$)} & \mc{$r_{\rm A}$} & \mc{$r_{\rm b}$} \\
\hline
DFOSC $R$-band  & 0.3703 & 0.0055 & 0.1129 & 0.0009 & 88.1 & 2.2 & 0.3328 & 0.0048 & 0.03755 & 0.00074 \\
DFOSC $I$-band  & 0.3766 & 0.0146 & 0.1118 & 0.0013 & 84.8 & 4.2 & 0.3387 & 0.0128 & 0.03788 & 0.00175 \\
GROND $g$-band  & 0.3734 & 0.0140 & 0.1183 & 0.0022 & 86.3 & 3.9 & 0.3339 & 0.0123 & 0.03949 & 0.00201 \\
GROND $r$-band  & 0.3753 & 0.0102 & 0.1150 & 0.0011 & 85.2 & 3.2 & 0.3366 & 0.0087 & 0.03870 & 0.00124 \\
GROND $i$-band  & 0.3667 & 0.0132 & 0.1091 & 0.0017 & 87.1 & 3.6 & 0.3307 & 0.0116 & 0.03606 & 0.00129 \\
GROND $z$-band  & 0.3661 & 0.0111 & 0.1106 & 0.0016 & 89.9 & 2.4 & 0.3297 & 0.0099 & 0.03645 & 0.00120 \\
CASLEO $R$-band & 0.3665 & 0.0203 & 0.1117 & 0.0055 & 89.6 & 4.5 & 0.3296 & 0.0167 & 0.03683 & 0.00316 \\
\hline
Final results&{\bf0.3712}&{\bf0.0040}&{\bf0.1127}&{\bf0.0009}&{\bf87.3}&{\bf1.2}&{\bf0.3335}&{\bf0.0035}&{\bf0.03754}&{\bf0.00049}\\
\hline
G14 & \mc{0.3725} & \erc{0.1093}{0.0019}{0.0017} & 86.3 & 2.7 & \erc{0.3358}{0.0111}{0.0055} & \mc{0.03670} \\
\hline \end{tabular} \end{table*}

We analysed our light curves using the {\sc jktebop}\footnote{{\sc jktebop} is written in {\sc fortran77} and the source code is available at {\tt http://www.astro.keele.ac.uk/jkt/codes/jktebop.html}} code \citep{Me++04mn} and the {\it Homogeneous Studies} methodology \citep[][and references therein]{Me12mn}. The light curves were divided up according to their passband (Bessell $R$ and $I$ for DFOSC and SDSS $griz$ for GROND) and each set was modelled together.

The model was parameterised by the fractional radii of the star and the planet ($r_{\rm A}$ and $r_{\rm b}$), which are the ratios between the true radii and the semimajor axis ($r_{\rm A,b} = \frac{R_{\rm A,b}}{a}$). The parameters of the fit were the sum and ratio of the fractional radii ($r_{\rm A} + r_{\rm b}$ and $k = \frac{r_{\rm b}}{r_{\rm A}}$), the orbital inclination ($i$), limb darkening coefficients, and a reference time of mid-transit. We assumed an orbital eccentricity of zero (G14) and the orbital period found in Section\,\ref{sec:porb}. We also fitted for the coefficients of polynomial functions of differential magnitude versus time \citep{Me+14mn}. One polynomial was used for each transit light curve, of the order given in Table\,\ref{tab:obslog}.

Limb darkening was incorporated using each of five laws \citep[see][]{Me08mn}, with the linear coefficients either fixed at theoretically predicted values\footnote{Theoretical limb darkening coefficients were obtained by bilinear interpolation in \Teff\ and \logg\ using the {\sc jktld} code available from: {\tt http://www.astro.keele.ac.uk/jkt/codes/jktld.html}} or included as fitted parameters. We did not calculate fits for both limb darkening coefficients in the four two-coefficient laws as they are very strongly correlated \citep{Me++07aa,Carter+08apj}. The nonlinear coefficients were instead perturbed by $\pm$0.1 on a flat distribution during the error analysis simulations, in order to account for imperfections in the theoretical values of the coefficients.

Error estimates for the fitted parameters were obtained in three ways. Two sets were obtained using residual-permutation and Monte Carlo simulations \citep{Me08mn} and the larger of the two was retained for each fitted parameter. We also ran solutions using the five different limb darkening laws, and increased the errorbar for each parameter to account for any disagreement between these five solutions. Tables of results for each light curve can be found in the Appendix and the best fits can be inspected in Fig.\,\ref{fig:lcfit}.

\subsection{Results}

For all light curves we found that the best solutions were obtained when the linear limb darkening coefficient was fitted and the nonlinear coefficient was fixed but perturbed. We found that there is a significant correlation between $i$ and $k$ for all light curves, which hinders the precision to which we can measure the photometric parameters. The best fit for the CASLEO and the GROND $z$-band data is a central transit ($i \approx 90^\circ$), but this does not have a significant effect on the value of $k$ measured from these data.

Table\,\ref{tab:lcfit} holds the measured parameters from each light curve. The final value for each parameter is the weighted mean of the values from the different light curves. We find a good agreement for all parameters except for $k$, which is in line with previous experience (see \citealt{Me12mn} and references therein). The \chir\ of the individual values of $k$ versus the weighted mean is 3.1, and the errorbar for the final value of $k$ in Table\,\ref{tab:lcfit} has been multiplied by $\sqrt{3.1}$ to force a \chir\ of unity. Our results agree with, but are significantly more precise than, those found by G14.


\section{Physical properties}                                                                                                      \label{sec:absdim}

\begin{table*} \caption{\label{tab:model} Derived physical properties of WASP-103. Quantities marked with a $^\star$ are significantly
affected by the spherical approximation used to model the light curves, and revised values are given at the base of the table.}
\begin{tabular}{l l l r@{\,$\pm$\,}c@{\,$\pm$\,}l r@{\,$\pm$\,}l} \hline
Quantity                & Symbol           & Unit  & \mcc{This work}                 & \mc{G14}                       \\
\hline
Stellar mass            & $M_{\rm A}$      & \Msun & 1.204    & 0.089    & 0.019     & \erc{1.220}{0.039}{0.036}      \\
Stellar radius          & $R_{\rm A}$      & \Rsun & 1.419    & 0.039    & 0.008     & \erc{1.436}{0.052}{0.031}      \\
Stellar surface gravity & $\log g_{\rm A}$ & \,cgs & 4.215    & 0.014    & 0.002     & \erc{4.22}{0.12}{0.05}         \\
Stellar density         & $\rho_{\rm A}$   & \psun & \mcc{$0.421 \pm 0.013$}         & \erc{0.414}{0.021}{0.039}      \\[2pt]
Planet mass             & $M_{\rm b}$      & \Mjup & 1.47     & 0.11     & 0.02      & 1.490 & 0.088                  \\
Planet radius$^\star$   & $R_{\rm b}$      & \Rjup & 1.554    & 0.044    & 0.008     & \erc{1.528}{0.073}{0.047}      \\
Planet surface gravity  & $g_{\rm b}$      & \mss  & \mcc{$15.12 \pm  0.93$}         & 15.7 & 1.4                     \\
Planet density$^\star$  & $\rho_{\rm b}$   & \pjup & 0.367    & 0.027    & 0.002     & \erc{0.415}{0.046}{0.053}      \\[2pt]
Equilibrium temperature & \Teq\            & K     & \mcc{$2495 \pm   66$}           & \erc{2508}{75}{70}             \\
Safronov number         & \safronov\       &       & 0.0311   & 0.0019   & 0.0002    & \mc{ }                         \\
Orbital semimajor axis  & $a$              & au    & 0.01978  & 0.00049  & 0.00010   & 0.01985 & 0.00021              \\
Age                     & $\tau$           & Gyr   & \ermcc{3.8}{2.1}{1.6}{0.3}{0.4} & \mc{3 to 5}                    \\
\hline
\multicolumn{6}{l}{{\it Planetary parameters corrected for asphericity:}} \\
Planet radius           &                  & \Rjup & \mcc{$1.603 \pm 0.052$} \\
Planet density          &                  & \pjup & \mcc{$0.335 \pm 0.025$} \\
\hline \end{tabular} \end{table*}

We have measured the physical properties of the WASP-103 system using the results from Section\,\ref{sec:lc}, five grids of predictions from theoretical models of stellar evolution \citep{Claret04aa,Demarque+04apjs,Pietrinferni+04apj,Vandenberg++06apjs,Dotter+08apjs}, and the spectroscopic properties of the host star. Theoretical models are needed to provide an additional constraint on the stellar properties as the system properties cannot be obtained from only measured quantities. The spectroscopic properties were obtained by G14 and comprise effective temperature ($\Teff = 6110 \pm 160$\,K), metallicity ($\FeH = 0.06 \pm 0.13$) and velocity amplitude ($K_{\rm A} = 271 \pm 15$\ms). The adopted set of physical constants is given in \cite{Me11mn}.

We first estimated the velocity amplitude of the {\em planet}, $K_{\rm b}$, and used this along with the measured $r_{\rm A}$, $r_{\rm b}$, $i$ and $K_{\rm A}$ to determine the physical properties of the system. $K_{\rm b}$ was then iteratively refined to find the best match between the measured $r_{\rm A}$ and the calculated $\frac{R_{\rm A}}{a}$, and the observed \Teff\ and that predicted by a theoretical model for the obtained stellar mass, radius and \FeH. This was done for a grid of ages from the zero-age main sequence to beyond the terminal-age main sequence for the star, in 0.01\,Gyr increments, and the overall best $K_{\rm b}$ was adopted. The statistical errors in the input quantities were propagated to the output quantities by a perturbation approach.

We ran the above analysis for each of the five grids of theoretical stellar models, yielding five different estimates of each output quantity. These were transformed into a single final result for each parameter by taking the unweighted mean of the five estimates and their statistical errors, plus an accompanying systematic error which gives the largest difference between the mean and individual values. The final results of this process are a set of physical properties for the WASP-103 system, each with a statistical error and a systematic error. The stellar density, planetary surface gravity and planetary equilibrium temperatures can be calculated without resorting to theoretical predictions \citep{SeagerMallen03apj,Me++07mn,Me10mn}, so do not have an associated systematic error.

\subsection{Results}

\begin{table} \centering
\caption{\label{tab:err} Detailed error budget for the calculation of the system properties
of WASP-103 from the photometric and spectroscopic parameters, and the Y$^2$ stellar
models. Each number in the table is the fractional contribution to the final uncertainty of an
output parameter from the errorbar of an input parameter. The final uncertainty for each output
parameter is the quadrature sum of the individual contributions from each input parameter.}
\begin{tabular}{l|cccccc} \hline
Output            & \multicolumn{6}{c}{Input parameter} \\
parameter         & $K_{\rm A}$ & $i$   & $r_{\rm A}$ & $r_{\rm b}$ & \Teff & \FeH  \\
\hline
Age               &    0.012    &       &    0.035    &             & 0.873 & 0.471 \\
$a$               &    0.030    &       &    0.020    &             & 0.797 & 0.601 \\[2pt]
$M_{\rm A}$       &    0.029    &       &    0.020    &             & 0.796 & 0.602 \\
$R_{\rm A}$       &    0.027    &       &    0.469    &             & 0.703 & 0.530 \\
$\log g_{\rm A}$  &    0.019    &       &    0.706    &             & 0.563 & 0.424 \\
$\rho_{\rm A}$    &    0.002    &       &    1.000    &             & 0.001 &       \\[2pt]
$M_{\rm b}$       &    0.809    & 0.014 &    0.012    &             & 0.466 & 0.352 \\
$R_{\rm b}$       &    0.025    &       &    0.017    &    0.544    & 0.668 & 0.504 \\
$g_{\rm b}$       &    0.901    & 0.016 &             &    0.434    &       &       \\
$\rho_{\rm b}$    &    0.772    & 0.014 &    0.006    &    0.564    & 0.232 & 0.174 \\
\hline \end{tabular} \end{table}

Our final results are given in Table\,\ref{tab:model} and have been added to TEPCat\footnote{TEPCat is The Transiting Extrasolar Planet Catalogue \citep{Me11mn} at: {\tt http://www.astro.keele.ac.uk/jkt/tepcat/}}. We find a good agreement between the five different model sets (Table\,A8). Some of the measured quantities, in particular the stellar and planetary mass, are still relatively uncertain. To investigate this we calculated a complete error budget for each output parameter, and show the results of this analysis in Table\,\ref{tab:err} when using the Y$^2$ theoretical stellar models \citep{Demarque+04apjs}. The error budgets for the other four model sets are similar.

The uncertainties in the physical properties of the planet are dominated by that in $K_{\rm A}$, followed by that in $r_{\rm b}$. The uncertainties in the stellar properties are dominated by those in \Teff\ and \FeH, followed by $r_{\rm A}$. Improvements in our understanding of the WASP-103 system would most easily be achieved by obtaining new spectra from which additional radial velocity measurements and improved \Teff\ and \FeH\ measurements could be obtained.

To illustrate the progress possible from further spectroscopic analysis, we reran the analysis but with smaller errorbars of $\pm$50\,K in \Teff\ and $\pm$0.05\,dex in \FeH. The precision in $M_{\rm A}$ changes from 0.091\Msun\ to 0.041\Msun. Similar improvements are seen for $a$, and smaller improvements for $R_{\rm A}$, $R_{\rm b}$ and $\rho_{\rm b}$. Augmenting this situation by adopting an errorbar of $\pm$5\ms\ in $K_{\rm A}$ changes the precision in $M_{\rm b}$ from 0.11\Mjup\ to 0.047\Mjup\ and yields further improvements for $R_{\rm b}$ and $\rho_{\rm b}$.

\subsection{Comparison with G14}

Table\,\ref{tab:model} also shows the parameter values found by G14, which are in good agreement with our results. Some of the errorbars, however, are smaller than those in the current work, despite the fact that G14 had much less observational data at their disposal. A possible reason for this discrepancy is the additional constraint used to obtain a determinate model for the system. We used each of five sets of theoretical model predictions, whilst G14 adopted a calibration of $M_{\rm A}$ as a function of $\rho_{\rm A}$, \Teff\ and \FeH\ based on semi-empirical results from the analysis of low-mass detached eclipsing binary (dEB) systems \citep{Torres++08apj,Enoch+10aa,Me09mn,Me11mn}. The dEB calibration suffers from an astrophysical scatter of the calibrating objects which is much greater than that of the precision to which the calibration function can be fitted \citep[see][]{Me11mn}. G14 accounted for the uncertainty in the calibration by perturbing the measured properties of the calibrators during their Markov \reff{chain} Monte Carlo analysis \citep{Gillon+13aa}. They therefore accounted for the observational uncertainties in the measured properties of the calibrators, but \reff{neglected} the astrophysical scatter.

There is supporting evidence for this interpretation of why our errorbars for some measurements are significantly larger than those found by G14. Our own implementation of the dEB calibration \citep{Me11mn} explicitly includes the astrophysical scatter and yields $M_{\rm A} = 1.29 \pm 0.11$\Msun, where the greatest contribution to the uncertainty is the scatter of the calibrators around the calibration function. \reff{G14 themselves found a value of $M_{\rm A} = 1.18 \pm 0.10$\Msun\ from an alternative approach (comparable to our main method) of using the CL\'ES theoretical models \citep{Scuflaire+08apss} as their additional constraint. This is much less precise than their default value of $M_{\rm A} = \er{1.220}{0.039}{0.036}$\Msun\ from the dEB calibration.} M.\ Gillon (private communication) confirms our interpretation of the situation.

\subsection{Correction for asphericity}

%

\begin{table} \begin{center}
\caption{\label{tab:sphere} Specification of the shape of WASP-103\,b obtained using Roche geometry.}
\begin{tabular}{llr@{\,$\pm$\,}l} \hline
Symbol & Description & \mc{Value}  \\
\hline
$R_{\rm sub}$ (\Rjup)               & Radius at substellar point  & 1.721 & 0.075 \\
$R_{\rm back}$ (\Rjup)              & Radius at antistellar point & 1.710 & 0.072 \\
$R_{\rm side}$ (\Rjup)              & Radius at sides             & 1.571 & 0.047 \\
$R_{\rm pole}$ (\Rjup)              & Radius at poles             & 1.537 & 0.043 \\
$R_{\rm cross}$ (\Rjup)             & Cross-sectional radius      & 1.554 & 0.045 \\
$R_{\rm mean}$ (\Rjup)              & Mean radius                 & 1.603 & 0.052 \\[3pt]
$f_{\rm RL}$                        & Roche lobe filling factor   & 0.584 & 0.033 \\[3pt]
$R_{\rm sub}/R_{\rm side}$          &                             & 1.095 & 0.017 \\
$R_{\rm sub}/R_{\rm pole}$          &                             & 1.120 & 0.020 \\
$R_{\rm side}/R_{\rm pole}$         &                             & 1.022 & 0.003 \\
$R_{\rm back}/R_{\rm sub}$          &                             & 0.994 & 0.002 \\
$R_{\rm mean}/R_{\rm cross}^{\ 3}$  & {\it density correction}    & 1.096 & 0.015 \\
\hline \end{tabular} \end{center}
\Rjup, the equatorial radius of Jupiter, is adopted to be 71\,492\,km.
\end{table}

As pointed out by \citet{Li+10nat} for the case of WASP-12\,b, some close-in extrasolar planets may have significant departures from spherical shape. \citet{Budaj11aj} calculated the Roche shapes of all transiting planets known at that time, as well as light curves and spectra taking into account the non-spherical shape. He found that WASP-19\,b and WASP-12\,b had the most significant tidal distortion of all known planets. The Roche model assumes that the object is rotating synchronously with the orbital period, there is a negligible orbital eccentricity, and that masses can be treated as point masses. The Roche shape has a characteristic pronounced expansion of the object towards the sub-stellar point, and a slightly less pronounced expansion towards the anti-stellar point. The radii on the side of the object are smaller, and the radii at the rotation poles are the smallest. \citet{Leconte++11aa} developed a model of tidally distorted planets which takes into account the tidally distorted mass distribution within the object assuming an
ellipsoidal shape. \citet{Burton+14apj} studied the consequences of the Roche shape on the measured densities of exoplanets.

The Roche shape of a planet is determined by the semi-major axis, mass ratio, and a value of the surface potential. Assuming the parameters found above ($a = 4.25 \pm 0.11$\Rsun, $M_{\rm A}/M_{\rm b} = 854 \pm 4$ and $R_{\rm b} = 1.554 \pm 0.45$\Rjup) one can estimate the tidally distorted Roche potential, i.e.\ the shape of the planet which would have the same cross-section during the transit as the one inferred from the observations under the assumption of a spherical planet. The shape of WASP-103\,b is described by the parameters $R_{\rm sub}$, $R_{\rm back}$, $R_{\rm side}$ and $R_{\rm pole}$ (see \citealt{Budaj11aj} for more details). The descriptions and values of these are given in Table\,\ref{tab:sphere}. The uncertainties in Table\,\ref{tab:sphere} are the quadrature addition of those due to each input parameter; they are dominated by the uncertainty in the radius of the planet.

The cross-sectional radius, $R_{\rm cross} = \sqrt{R_{\rm side}R_{\rm pole}}$, is the radius of the circle with the same cross-section as the Roche surface during the transit. $R_{\rm cross}$ is the quantity measured from transit light curves using spherical-approximation codes such as {\sc jktebop}. $R_{\rm mean}$ is the radius of a sphere with the same volume as that enclosed by the Roche surface.

Table\,\ref{tab:sphere} also gives ratios between $R_{\rm sub}$, $R_{\rm back}$, $R_{\rm side}$ and $R_{\rm pole}$. Moderate changes in the planetary radius lead to very small changes in the ratios. In the case that future analyses yield a revised planetary radius, these ratios can therefore be used to rescale the values of $R_{\rm sub}$, $R_{\rm back}$, $R_{\rm side}$ and $R_{\rm pole}$ appropriately. In particular, the quantity $R_{\rm mean}/R_{\rm cross}^{\ 3}$ is the correction which must be applied to the density measured in the spherical approximation to convert it to the density obtained using Roche geometry.

WASP-103\,b \reff{has a Roche lobe filling factor ($f_{\rm RL}$) of 0.58, where $f_{\rm RL}$ is defined to be the radius of the planet at the substellar point relative to the radius of the L1 point. The planet} is therefore well away from Roche-lobe overflow but is significantly distorted. The above analysis provides corrections to the properties measured in the spherical approximation. The planetary radius increases by 2.2\% to $R_{\rm b} = 1.603 \pm 0.052$\Rjup, and its density falls by 9.6\% to $\rho_{\rm b} = 0.335 \pm 0.025$\pjup. These revised values include the uncertainty in the correction for asphericity and are included in Table\,\ref{tab:model}. These departures from sphericity mean WASP-103\,b is one of the three most distorted planets known, alongside WASP-19\,b and WASP-12\,b.


\section{Variation of radius with wavelength}                                                                                       \label{sec:rayleigh}

%


\begin{figure} \includegraphics[width=\columnwidth,angle=0]{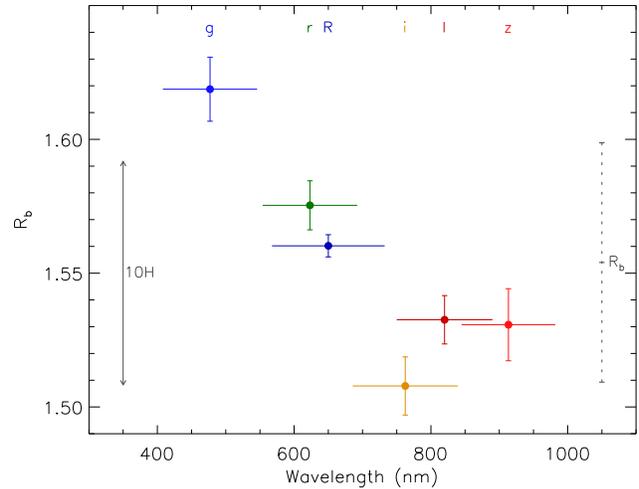}
\caption{\label{fig:rvary} Measured planetary radius ($R_{\rm b}$) as a function of the
central wavelength of the passbands used for different light curves. The datapoints show
the $R_{\rm b}$ measured from each light curve. The vertical errorbars show the relative
uncertainty in $R_{\rm b}$ (i.e.\ neglecting the common sources of error) and the horizontal
errorbars indicate the FWHM of the passband. The datapoints are colour-coded consistently
with Figs.\ \ref{fig:lc:dk} and \ref{fig:lc:grond} and the passbands are labelled at the
top of the figure. The \reff{dotted} grey line to the right of the figure shows the measured
value of $R_{\rm b}$ from Table\,\ref{tab:model}, which includes all sources of uncertainty. The
\reff{solid} grey line to the left of the figure shows how big ten pressure scale heights is.} \end{figure}

\begin{table} \centering
\caption{\label{tab:rb} Values of $r_{\rm b}$ for each of the light curves as plotted in
Fig.\,\ref{fig:rvary}. Note that the errorbars in this table exclude all common sources
of uncertainty in $r_{\rm b}$ so should only be used to compare different values of
$r_{\rm b}(\lambda)$.}
\begin{tabular}{lllr@{\,$\pm$\,}l} \hline
Passband & Central         & FWHM  & \mc{$r_{\rm b}$}  \\
         & wavelength (nm) & (nm)  & \mc{ }            \\
\hline
$g$      & \reff{477.0}    & 137.9 & 0.03911 & 0.00029 \\
$r$      & \reff{623.1}    & 138.2 & 0.03806 & 0.00022 \\
$R$      & \reff{658.9}    & 164.7 & 0.03770 & 0.00010 \\
$i$      & \reff{762.5}    & 153.5 & 0.03643 & 0.00026 \\
$I$      & \reff{820.0}    & 140.0 & 0.03703 & 0.00022 \\
$z$      & \reff{913.4}    & 137.0 & 0.03698 & 0.00033 \\
\hline \end{tabular} \end{table}

If a planet has an extended atmosphere, then a variation of opacity with wavelength will cause a variation of the measured planetary radius with wavelength. The light curve solutions (Table\,\ref{tab:lcfit}) show a dependence between the measured value of $k$ and the central wavelength of the passband used, in that larger $k$ values occur at bluer wavelengths. This implies a larger planetary radius in the blue, which might be due to Rayleigh scattering from a high-altitude atmospheric haze \citep[e.g.][]{Pont+08mn,Sing+11mn,Pont+13mn}.

We followed the approach of \citet{Me+12mn2} to tease out this signal from our light curves. We modelled each dataset with the parameters $r_{\rm A}$ and $i$ fixed at the final values in Table\,\ref{tab:lcfit}, but still fitting for $T_0$, $r_{\rm b}$, the linear limb darkening coefficient and the polynomial coefficients. \reff{We did not consider solutions with both limb darkening coefficients fixed, as they had a significantly poorer fit, or with both fitted, as this resulted in unphysical values of the coefficients for most of the light curves.} This \reff{process} yielded a value of $r_{\rm b}$ for each light curve with all common sources of \reff{uncertainty} removed from the errorbars (Table\,\ref{tab:rb}), which we then converted to $R_{\rm b}$ using the semimajor axis from Table\,\ref{tab:model}. We discluded the CASLEO data from this analysis due to the low precision of the $r_{\rm b}$ it gave.

Fig.\,\ref{fig:rvary} shows the $R_{\rm b}$ values found from the individual light curves as a function of wavelength. The value of $R_{\rm b}$ from Table\,\ref{tab:model} is indicated for context. We calculated the atmospheric pressure scale height, $H$, of WASP-103\,b using this formula \citep[e.g.][]{DepaterLissauer01book}:
\begin{equation}
H = \frac{ k_{\rm B} \Teq }{ \mu g_{\rm b} }
\end{equation}
where $k_{\rm B}$ is Botzmann's constant and $\mu$ is the mean molecular weight in the atmosphere. We adopted $\mu = 2.3$ following \citet{DewitSeager13sci}, and the other parameters were taken from Table\,\ref{tab:model}. This yielded $H = 597$\,km $= 0.00834$\Rjup. The relative errors on our individual $R_{\rm b}$ values are therefore in the region of few pressure scale heights, and the total variation we find between the $g$ and $i$ bands is $13.3H$. For comparison, \citet[][their fig.\,14]{Sing+11mn} found a variation of $6H$ between 330\,nm and 1\,$\mu$m in transmission spectra of HD\,189733\,b. Similar or larger effects have been noted in transmission photometry of HAT-P-5 \citep{Me+12mn2}, GJ\,3470 \citep{Nascimbeni+13aa2} and Qatar-2 \citep{Mancini+14mn}.

Is this variation with wavelength plausible? To examine this we turned to the MassSpec concept proposed by \citet{DewitSeager13sci}. The atmospheric scale height depends on surface gravity and thus the planet mass:
\begin{equation}
M_{\rm b} = \frac{ k_{\rm B} \Teq R_{\rm b}^2 }{ \mu G H }
\end{equation}
where $G$ is the gravitational constant. The variation of the measured radius with wavelength due to Rayleigh scattering depends on the atmospheric scale height under the assumption of a power law relation between the wavelength and cross-section of the scattering species. Rayleigh scattering corresponds to a power law coefficient of $\alpha = -4$ \citep{Lecavelier+08aa} where
\begin{equation}
\alpha H = \frac{{\rm d}R_{\rm b}(\lambda)}{{\rm d}\ln\lambda}
\end{equation}
which yields the equation
\begin{equation}
M_{\rm b} = - \frac{ \alpha k_{\rm B} \Teq [ R_{\rm b}(\lambda) ]^2 }{ \mu G \frac{{\rm d}R_{\rm b}(\lambda)}{{\rm d}\ln\lambda} }
\end{equation}

We applied MassSpec to our $R_{\rm b}(\lambda)$ values for WASP-103\,b. The slope of $R_{\rm b}$ versus $\ln\lambda$ is detected to a significance of $7.3\sigma$ and corresponds to a planet mass of $0.31 \pm 0.05$\Mjup. The slope is robustly detected, but gives a planet mass much lower than the mass of $M_{\rm b} = 1.49 \pm 0.11$\Mjup\ found in Section\,\ref{sec:absdim}. The gradient of the slope is greater than it should be under the scenario outlined above. We can equalise the two mass measurements by adopting a stronger power law with a coefficient of $\alpha = 19.0 \pm 1.5$, which is extremely large. We conclude that our data are not consistent with Rayleigh scattering so are either affected by additional physical processes or are returning spurious results.

The presence of unocculted starspots on the visible disc of the star could cause a trend in the measured planetary radius similar to what we see for WASP-103. Unocculted spots cause an overestimate of the ratio of the radii \citep[e.g.][]{Czesla+09aa,Ballerini+12aa,Oshagh+13aa}, and are cooler than the surrounding photosphere so have a greater effect in the blue. They therefore bias planetary radius measurements to higher values, and do so more strongly at bluer wavelengths. Occulted plage can cause an analagous effect \citep{Oshagh+14aa}, but we are aware of only circumstantial evidence for plage of the necessary brightness and extent in planet host stars.

The presence of starspots has not been observed on WASP-103\,A, which at $\Teff = 6110$\,K is too hot to suffer major spot activity. G14 found no evidence for spot-induced rotational modulation down to a limiting amplitude of 3\,mmag, and our light curves show no features attributable to occultations of a starspot by the planet. This is therefore an unlikely explanation for the strong correlation between $R_{\rm b}$ and wavelength. Further investigation of this effect requires data with a greater spectral coverage and/or resolution.


\section{Summary and conclusions}                                                                                                 \label{sec:summary}

The recently-discovered planetary system WASP-103 is well suited to detailed analysis due to its short orbital period and the brightness of the host star. These analyses include the investigation of tidal effects, the determination of high-precision physical properties, and the investigation of the atmospheric properties of the planet. We have obtained 17 new transit light curves which we use to further our understanding in all three areas.

The extremely short orbital period of the WASP-103 system makes it a strong candidate for the detection of tidally-induced orbital decay \citep{Birkby+14mn}. \reff{Detecting this effect could yield a measurement of the tidal quality factor for the host star, which is vital for assessing the strength of tidal effects such as orbital circularisation, and for predicting the ultimate fate of hot Jupiters.} The prime limitation in \reff{attempts to observe this effect} is that the strength of the signal, and therefore the length of the observational program needed to detect it, is unknown. Our high-precision light curves improve the measurement of the time of midpoint of a transit at the reference epoch from 67.4\,s (G14) to 4.8\,s (this work). There are currently no indications of a change in orbital period, but these effects are expected to take of order a decade to become apparent. Our work establishes a high-precision \reff{transit timing} at the reference epoch against which future observations can be measured.

We modelled our light curves with the {\sc jktebop} code following the {\it Homogeneous Studies} methodology in order to measure high-precision photometric parameters of the system. These were combined with published spectroscopic results and with five sets of theoretical stellar models in order to determine the physical properties of the system to high precision and with robust error estimates. We present an error budget which shows that more precise measurements of the \Teff\ and \FeH\ of the host star would be an effective way of further improving these results. A high-resolution {\it Lucky Imaging} observation shows no evidence for the presence of faint stars at small (but non-zero) angular separations from WASP-103, which might have contaminated the flux from the system and thus caused us to underestimate the radius of the planet.

The short orbital period of the planet means it is extremely close to its host star: its orbital separation of 0.01978\,au corresponds to only $3.0R_{\rm A}$. This distorts the planet from a spherical shape, and causes an underestimate of its radius when light curves are modelled in the spherical approximation. We determined the planetary shape using Roche geometry \citep{Budaj11aj} and utilized these results to correct its measured radius and mean density for the effects of asphericity.

Our light curves were taken in six passbands spanning much of the optical wavelength region. There is a trend towards finding a larger planetary radius at bluer wavelengths, at a statistical significance of $7.3\sigma$. We used the MassSpec concept \citep{DewitSeager13sci} to convert this into a measurement of the planetary mass under the assumption that the slope is caused by Rayleigh scattering. The resulting mass is too small by a factor of five, implying that Rayleigh scattering is not the main culprit for the observed variation of radius with wavelength.

We recommend that further work on the WASP-103 system includes a detailed spectral analysis for the host star, transit depth measurements in the optical and infrared with a higher spectral resolution than achieved here, and occultation depth measurements to determine the thermal emission of the planet and thus constrain its atmospheric energy budget. Long-term monitoring of its times of transit is also necessary in order to detect the predicted orbital decay due to tidal effects. \reff{Finally, the system is a good candidate for observing the Rossiter-McLaughlin effect, due to the substantial rotational velocity of the star ($v \sin i = 10.6 \pm 0.9$\kms; G14).}


\section*{Acknowledgements}

The operation of the Danish 1.54m telescope is financed by a grant to UGJ from the Danish Natural Science Research Council (FNU).
This paper incorporates observations collected using the Gamma Ray Burst Optical and Near-Infrared Detector (GROND) instrument at the MPG 2.2\,m telescope located at ESO La Silla, Chile, program 093.A-9007(A). GROND was built by the high-energy group of MPE in collaboration with the LSW Tautenburg and ESO, and is operated as a PI-instrument at the MPG 2.2\,m telescope. 
We thank Mike Gillon for helpful discussions.
The reduced light curves presented in this work will be made available at the CDS ({\tt http://vizier.u-strasbg.fr/}) and at {\tt http://www.astro.keele.ac.uk/$\sim$jkt/}.
J\,Southworth acknowledges financial support from STFC in the form of an Advanced Fellowship.
JB acknowledges funding by the Australian Research Council Discovery Project Grant DP120101792.
Funding for the Stellar Astrophysics Centre is provided by The Danish National Research Foundation (Grant agreement no.\ DNRF106).
TH is supported by a Sapere Aude Starting Grant from The Danish Council for Independent Research.
This publication was supported by grant NPRP X-019-1-006 from Qatar National Research Fund (a member of Qatar Foundation).
TCH is supported by the Korea Astronomy \& Space Science Institute travel grant \#2014-1-400-06.
CS received funding from the European Union Seventh Framework Programme (FP7/2007-2013) under grant agreement no.\ 268421.
%
%
%
%
%
%
OW (FNRS research fellow) and J\,Surdej acknowledge support from the Communaut\'e fran\c{c}aise de Belgique - Actions de recherche concert\'ees - Acad\'emie Wallonie-Europe.
The following internet-based resources were used in research for this paper: the ESO Digitized Sky Survey; the NASA Astrophysics Data System; the SIMBAD database and VizieR catalogue access tool operated at CDS, Strasbourg, France; and the ar$\chi$iv scientific paper preprint service operated by Cornell University.

\bibliographystyle{mn_new}

\end{document}